# Auto-Documenation for Software Development

AI-Assisted Programming

Thomas Zheng, Jeff Shaw, Sergey Kozlov
Luddite Labs, Inc.
San Diego, USA
Thomas.Zheng@ludditelabs.io
Jeff.Shaw@ludditelabs.io
Sergey.Kozlov@ludditelabs.io

*Abstract*—Software documentation is an essential but labor intensive task that often requires a dedicated team of developers to ensure coverage and accuracy. Good documentation will help shorten the development cycle and improve the overall team efficiency as well as maintainability. In today's crowd-driven development environment, good documentation can go a long way in building a developer community from scratch. To that end, we took the first steps in building a tool called Autodoc that can assist software developers in writing better documentation faster. Autodoc goes beyond traditional boilerplate template generation. Our integrated tool uses Deep Learning methods to construct a semantic understanding of the code. Just like machine translation in natural languages, Autodoc can translate snippets of code to comments, and insert them as short summaries inside the *docstring*. We also demonstrate the integration of Autodoc as an IDE plugin as well as a web hook from within software hosting platforms when submitting auto-documented code to user's Git repository.

## I. INTRODUCTION

Modern software development is a distributed team effort. Software hosting platforms such as Github and Bitbucket enable any team of arbitrary size to join effort in developing quality software. This social element makes documenting what one has contributed an essential component in the life cycle of software process. The benefits of good documentation in both coverage and accuracy grow exponentially as the team gets larger and larger. For example, in a crowd-driven development environment good documentation is a key factor in driving adoption and attracting active users to make productive contributions to the software repository.

### A. Cost of Documentation

There is an unspoken tendency for programmers to postpone documentation. Documentation can be labor intensive but it does not directly contribute to the functionality of a program. Furthermore, when the code changes, the documentation may change as well. This dilemma motivates a developer to defer the documentation until the end of the project when resources are even more constrained and there is even less incentive for a documentation effort. As a result, the documentation effort falls short of its goal of making the code more usable and maintainable.

### B. Tools for Boilerplate Documentation

Many tools exist today to help programmers build quality documentation templates. Doxygen, PanDoc, PerlDoc, PyDoc are just a few . Most of tools support the major development platforms and the most widely used programming languages such as C/C++, Java, Javascript and Python. Table I and II below are small snapshots of the current state of the art in boilerplate type documentation with more details found in [1].

TABLE I.     DOCUMENTATION TOOLS AND OS SUPPORT [1]

|  | Windows | OSX | Linux | Software License |
|---|---|---|---|---|
| **Doxygen** | X | X | X | GPL |
| **JavaDoc** | X | X | X | GPL |
| **HeaderDoc** | - | X | X | APSL |
| **Sphinx** | X | X | X | BSD |

TABLE II.     DOCUMENTATION TOOLS AND PROGRAMMING LANGUAGE SUPPORT [1]

|  | C/C++ | Java | JavaScript | Python |
|---|---|---|---|---|
| **Doxygen** | X | - | - | - |
| **JavaDoc** | - | X | - | - |
| **HeaderDoc** | X | X | X | X |
| **Sphinx** | X | with Plugin | X | X |

### C. From Code To Comment

The code-to-comment problem is similar to the machine translation problem for natural languages [2][3]. Language translation transforms a sequence of symbols or sentences from one language to another. For software documentation, we would like to translate code into a plain English description of its functionality so that others can understand what the code is doing. The state of art in machine translation today is using a sequence based neural network called Recurrent Neural Network (RNN). RNN has proven to be very powerful tool for translating between natural languages. An example of how it can be used is shown in Fig. 1. Others have tried to adopt the same architecture for translating code to comments with some success [4]. The key to adopting this new technology in the software development process is the flexibility and ease of upgrading to newer and better neural network solutions.

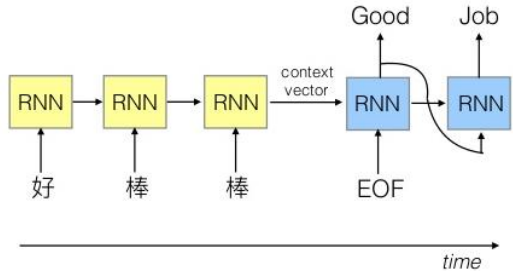

Figure 1. Example of a machine translation system using an RNN

## II. AUTODOC

The Autodoc tool was motivated by a desire to make our own Luddite Labs software development efforts more efficient. We wanted our developers to focus on programming and have Autodoc remove much of the burden of writing documentation without sacrificing quality. To make code documentation more accessible, we chose to integrate the Autodoc tool into the software development process with an emphasis on the ease of use. We experimented with two areas of deployment: Integration into an IDE as a plug-in and integration into the software hosting platform via web hooks.

In the following sections, we will describe the overall architecture of our Autodoc engine, the CTC (code-to-comment) generator block and its deployment channels that we have developed to enable others to write better software through automatically documenting their code.

### A. Autodoc Engine

Most popular programming languages use the concept of *docstring* as a literal string, specified in the source code, that is used to comment or otherwise document a segment of code. Autodoc, in a nutshell, is a post processor for a precompiled code. By manipulating the Abstract Syntax Tree (AST) we can insert docstrings into the code automatically. Fig. 2 below shows a high-level system architecture for the Autodoc tool. In the first step (Box 1), we use Sphinx and its various extension configurations to support various documentation templates. We settled on using Sphinx because it has a favorable software license and it supports the ~~three~~ four most-widely used programming languages: C/C++, Java, Javascript and Python. In the step (Box 2), we parse the source code to extract an AST data structure that is used to detect and extract the docstring components from the code including the location of docstring in the code. If a docstring field is missing, we will then synthesize a new docstring and insert it into the code.

Once the docstrings are extracted, the Autodoc tool will parse the docstring into another AST specifically for the docstring (Box 3). The Autodoc engine will then analyze the existing docstring, and create fixes and updates if needed. A separate engine for code-to-comment (CTC) generation is used to insert a summary into the docstring (Box 5), which we will describe in the following subsection. Finally, a newly created or properly updated docstring is reinserted into the code block with logic and formatting untouched (Box 4).

### B. Code-to-Summary Generator

As discussed above, code summarization or code-to-comment generation is still improving and we need to adopt a framework to keep up with the pace of upgrades and changes. As a result, we adopted a server-client model to accommodate future developments. Specifically, we put the CTC engine inside a Docker container[1] to create a self-contained virtual environment with virtually allocated resources. This 2-layer separation affords us the flexibility in both deploying a fast-evolving CTC engine and manage the work load of multiple clients. For experimentation, we used an existing CTC generator [4] available on Github[2].

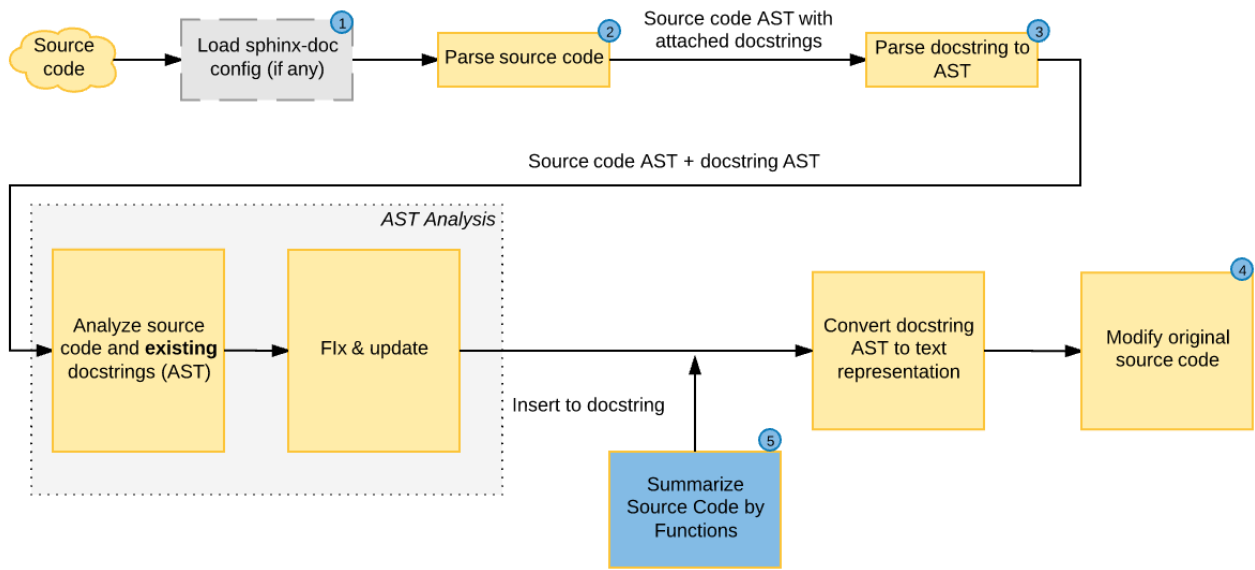

Figure 2 System Architecture for Autodoc engine

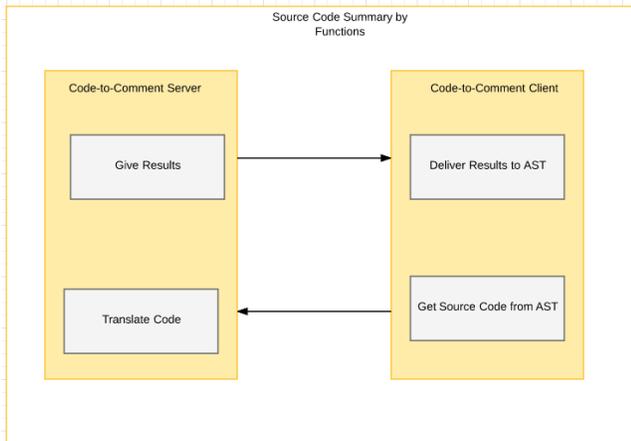

Figure 3: CTC engine architecture

## C. Integration into Software Hosting Platform

One way to deploy our Autodoc tool is to integrate it into the software repository hosting services such as Bitbucket and Github via web hooks. Once the user logs in to our Autodoc platform by their Bitbucket and/or Github account, the user can grant Autodoc access to one or more repositories. For the repositories which Autodoc can access, it will run the Autodoc engine for each commit made to the repository and put the change in a separate branch whose name is inspired by the existing branch. For example, if user recently committed a change on a master branch and Autodoc added a docstring or corrected other docstrings to the code, it will make those changes on top of the last commit on master and make them available on a master-autodoc branch (see Fig. 4). The user can merge the two branches manually or a pull request can be issued.

Figure 4: Branches generated by Autodoc

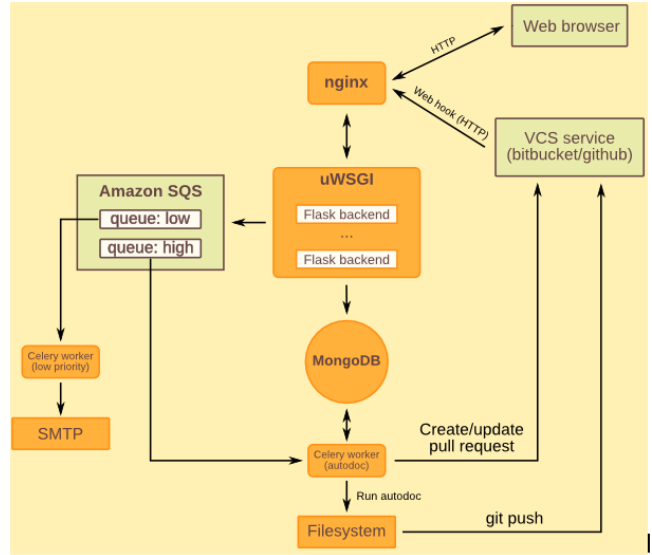

Figure 5: System architecture for Web Deployment

We use a MongoDB, a NO-SQL database for record keeping. To avoid complex configuration of shards in MongoDB, we started with single server solution. Another main component in this platform is the task-queue for parallel execution in the backend, for which we use SQS from AWS and Celery[3]. They are both widely used for building a distributed task queues via messaging a broker. In our case, we have two event queues:

- *high* – these are for events from the git repos. The main logic will be executed in this queue, so we need multiple workers here.
- *low* – these are for low priority tasks such as email sending, cleanup operations, stats aggregation.

To scale up, we can add more celery workers on queues. For better performance we add workers on different servers since each worker task consumes IO/CPU resources and increases the processing load on the queue listener. The concurrency of each worker can be controlled through the allocation of the available CPUs.

The web interface allows user to log into the Autodoc by using their Github or Bitbucket account. Upon login, the user can select the repositories to grant access for Autodoc (Fig. 6).

---

[1] See https://www.docker.com/what-docker
[2] Source: https://github.com/thaije/code-to-comment

[3] For celery: http://www.celeryproject.org/, For SQS: https://aws.amazon.com/sqs/

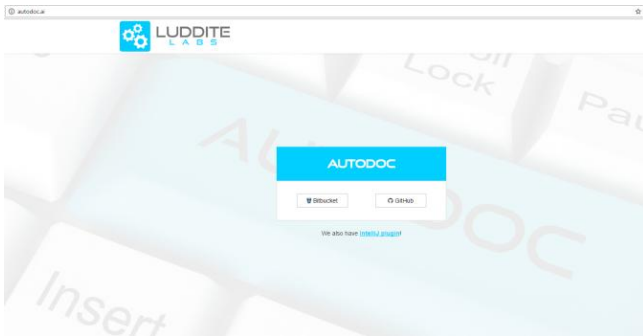

(a)

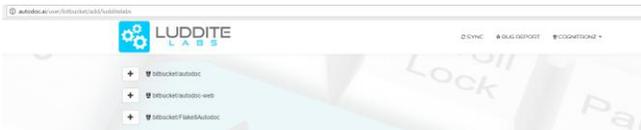

(b)

Figure 6: Web interface for the Autodoc platform. (a) the login page. (b) user can select the repositories in her account.

### D. Integrated Plugin in IDE

Another productive way of integrating the Autodoc tool is to integrate it into the IDE. At Luddite Labs we use PyCharm[4] and other IDEs developed by Jetbrains, so we developed a plugin for it. Fig. 7 shows the integrated plugin after installation on the Pycharm IDE. The Autodoc plugin can be downloaded from Pycharm's plugin page as shown in Fig. 8.

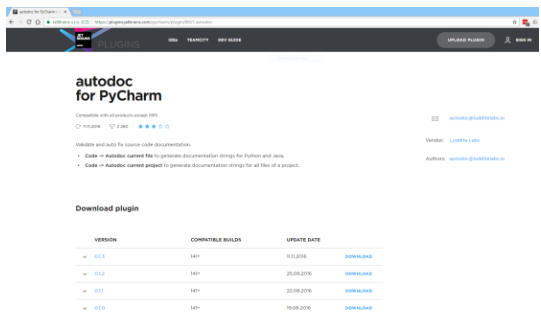

Figure 8: Pycharm plugin download for Autodoc

### III. CONCLUSION

Code documentation is an essential part of software development process but often neglected resulting in the accrual of large cumulative cost on an organization's software maintenance and growth. We discussed the design and development of an Autodoc tool which is being used by Luddite Labs and others to automatically generate documentation for written software. We also highlighted the integration of our Autodoc tool in both an IDE and a platform environment. Furthermore, we created a flexible framework for upgrading our code-to-comment generator so that we can quickly integrate a fast-evolving CTC engine.

---

[4] See https://www.jetbrains.com/pycharm/

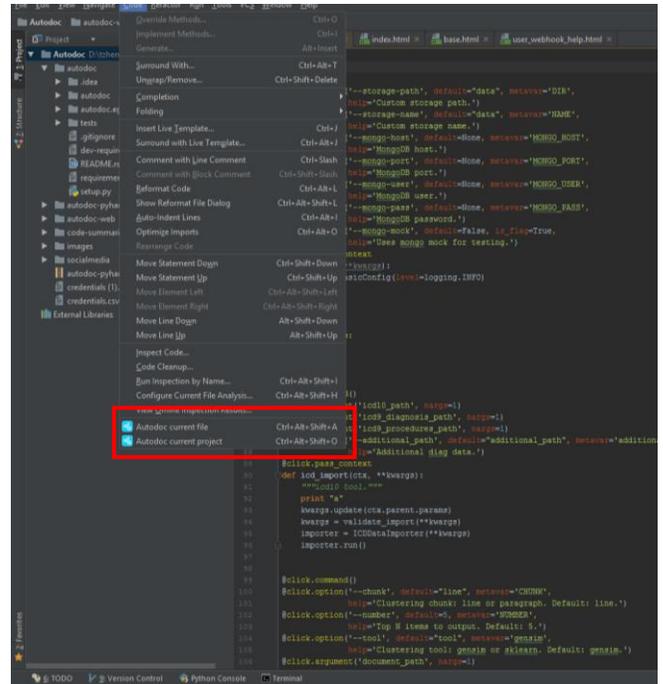

Figure 7. Autodoc plugin in Pycharm